\documentclass{aa}  

\usepackage{graphicx}
\usepackage{txfonts}
\usepackage[dvipsnames,table,xcdraw]{xcolor}
\usepackage[utf8]{inputenc}
\usepackage{xspace}
\usepackage{natbib}
\usepackage{listings}
\usepackage{multirow}
\usepackage{array}
\usepackage{amsmath}
\usepackage{etoolbox}
\usepackage{xtab}
\usepackage[normalem]{ulem}

\graphicspath{{figs/}}

\definecolor{codegreen}{rgb}{0,0.6,0}
\definecolor{codegray}{rgb}{0.5,0.5,0.5}
\definecolor{codepurple}{rgb}{0.58,0,0.82}
\definecolor{backcolour}{rgb}{0.95,0.95,0.92}

\lstdefinestyle{mystyle}{
    backgroundcolor=\color{backcolour},   
    commentstyle=\color{codegreen},
    keywordstyle=\color{magenta},
    numberstyle=\tiny\color{codegray},
    stringstyle=\color{codepurple},
    basicstyle=\ttfamily\footnotesize,
    breakatwhitespace=false,         
    breaklines=true,                 
    captionpos=b,                    
    keepspaces=true,                 
    numbers=none,                    
    numbersep=5pt,                  
    showspaces=false,                
    showstringspaces=false,
    showtabs=false,                  
    tabsize=2
}
\bibpunct{(}{)}{;}{a}{}{,} 
\usepackage{hyperref} 
\hypersetup{colorlinks=true, linktocpage=true, pdfstartpage=3,
            pdfstartview=FitV, breaklinks=true, pdfpagemode=UseNone,
            pageanchor=true, pdfpagemode=UseOutlines, plainpages=false,
            bookmarksnumbered, bookmarksopen=true, bookmarksopenlevel=1,
            hypertexnames=true, pdfhighlight=/O, urlcolor=RoyalBlue,
            linkcolor=RoyalBlue, citecolor=RoyalBlue}
            
\usepackage{catoptions}
\makeatletter

\def\Autoref#1{%
  \begingroup
  \edef\reserved@a{\cpttrimspaces{#1}}%
  \ifcsndefTF{r@#1}{%
    \xaftercsname{\expandafter\testreftype\@fourthoffive}
      {r@\reserved@a}.\\{#1}%
  }{%
    \ref{#1}%
  }%
  \endgroup
}
\def\testreftype#1.#2\\#3{%
  \ifcsndefTF{#1autorefname}{%
    \def\reserved@a##1##2\@nil{%
      \uppercase{\def\ref@name{##1}}%
      \csn@edef{#1autorefname}{\ref@name##2}%
      \autoref{#3}%
    }%
    \reserved@a#1\@nil
  }{%
    \autoref{#3}%
  }%
}
\makeatother

\newcommand{\source}[1]{\textsuperscript{\color{blue} [citation needed]}\xspace}

\newcommand{\bandu}{\ensuremath{u}\xspace}
\newcommand{\bandg}{\ensuremath{g}\xspace}
\newcommand{\bandr}{\ensuremath{r}\xspace}
\newcommand{\bandi}{\ensuremath{i}\xspace}
\newcommand{\bandz}{\ensuremath{z}\xspace}

\newcommand{\skybot}{\texttt{SkyBoT}\xspace}

\newcommand{\numb}[1]{\textcolor{orange}{#1}}
\renewcommand{\numb}[1]{#1}  

\renewcommand{\arcsec}{\ensuremath{^{\prime\prime}}\xspace}
\renewcommand{\arcmin}{\ensuremath{^{\prime}}\xspace}

\begin{document} 

\newcommand\mpcmoc{550,000\xspace}   
\newcommand\mpcnow{960,000\xspace}   
\newcommand\nummoc{471,569\xspace}    
\newcommand\nummocid{211,138\xspace}  
\newcommand\nummociduniq{100,322\xspace} 
\newcommand\numsvoid{57,646\xspace}  
\newcommand\numsvoiduniq{36,730\xspace} 

\newcommand\sdssDRFields{938,046\xspace} 
\newcommand\sdssStripeFields{746,919\xspace} 
\newcommand\sdssStripeUniqueFields{586,983\xspace} 

\newcommand\sdssDRfull{4,804,003\xspace} 
\newcommand\sdssSTRIPfull{4,071,153\xspace} 

\newcommand\ssops{3,238,202\xspace} 
\newcommand\ssogaia{3,238,135\xspace} 

\newcommand\ssoGroupsNumber{1,268,698\xspace} 
\newcommand\ssoAloneObserv{236,913\xspace} 

\newcommand\ssototprev{1,257,408\xspace} 
\newcommand\ssotot{1,542,522\xspace} 
\newcommand\ssoknown{1,036,322\xspace} 
\newcommand\ssouniq{379,714\xspace} 
\newcommand\ssounknown{506,200\xspace} 

\newcommand\skytot{812,573\xspace} 

\newcommand\ssoslow{216,348\xspace}
\newcommand\ssoslowbad{53,980\xspace}
\newcommand\knownslow{162,368\xspace}
\newcommand\fastnumber{9,134\xspace}
\newcommand\ssoNotInMoving{285,114\xspace}

\newcommand\ssoInADRfour{399,502\xspace}
\newcommand\ADRfourInPSOne{58,381\xspace}
\newcommand\ADRfourRemains{13,686\xspace}






\title{A million asteroid observations in the Sloan Digital Sky Survey}

\author{Alexey V. Sergeyev\inst{\ref{i:oca}} \and
          Benoit Carry\inst{\ref{i:oca}}}
\institute{Université Côte d'Azur, Observatoire de la Côte d'Azur, CNRS,
           Laboratoire Lagrange, France\\
           \email{alexey.sergeyev@oca.eu; benoit.carry@oca.eu}
           \label{i:oca}}
\date{\dots / \dots}

\abstract
{The populations of small bodies of the Solar System (asteroids, comets, Kuiper-Belt objects) are used to 
   constrain the origin and evolution of the Solar System. Both their orbital distribution and composition distribution
   are required to track the dynamical pathway from their regions of formation to their current locations.}
   {We aim at increasing the sample of Solar System objects (SSOs)  that have multi-filter photometry and compositional taxonomy.}
   {We search for moving objects in the archive of the Sloan Digital Sky Survey (SDSS).
   We attempt at maximizing the number of detections by using loose constraints on the extraction.
   We then apply a suite of filters to remove false-positive detections (stars or galaxies) and mark out spurious photometry and astrometry.}
   {We release a catalog of
   \numb{\ssotot} entries, consisting of 
   \numb{\ssoknown} observations of
   \numb{\ssouniq} known and unique SSOs together with
   \numb{\ssounknown} observations of moving sources not linked with any known SSOs.
   The catalog completeness is estimated to 
   be about \numb{95}\%
   and the purity to be above
   \numb{95}\%
   for known SSOs.}


\maketitle
%
\section{Introduction}

  \indent Considered as nuisances for the trails they imprint
  on photographic plates, asteroids (i.e., the ``vermin of the sky'') 
  are prime trackers of the dynamical events that shaped our Solar System, in particular migrations of the giant planets  \citep{2015-AsteroidsIV-Morbidelli, 2020MNRAS.492L..56C}.
  While their orbital distribution has been widely used, the distribution
  of their compositions provides further constraints on
  the models that describe the evolution of our Solar System 
  \citep{2009-Nature-460-Levison, 2016-AJ-152-Vokrouhlicky, 2017SciA....3E1138R}.

  \indent In this context, it is critical to understand their composition in detail and their location as well as the timing of formation.
  This is achieved through the laboratory
  study of meteorites \citep{2006-Nature-439-Bottke,
    2014ApJ...791..120V, 2018-ApJ-854-Scott},
  of which the near-Earth asteroids (NEAs) are the progenitors
  \citep{2018-Icarus-311-Granvik, 2018-Icarus-312-Granvik},
  and the compositional mapping of asteroids
  in the main belt (MB) between Mars and Jupiter
  \citep{2014Natur.505..629D}.

  \indent The composition, and more broadly the classification, of asteroids
  has been based for decades on multi-filter photometry in the visible
  \citep[e.g.,][]{1985-Icarus-61-Zellner}. With the advent of charge coupled devices (CCDs) 
  in the 1990s,
  multi-filter photometry moved toward low-resolution spectroscopy in the visible
  \citep{1995Icar..115....1X,2002Icar..158..146B} and was later extended to the near-infrared
  \citep{2009Icar..202..160D}.

  \indent After three decades of targeted observations, spectra
  (visible, near-infrared, or covering both 
  wave ranges) have been collected for about 9,000 asteroids 
  \citep[albeit disseminated over a myriad of articles, e.g.,][]{
          2002-Icarus-158-BusI, 2004Icar..172..179L, 2014Icar..233..163F},
  including around one thousand NEAs
  \citep[mainly from the NEOSHIELD2, MANOS, and MITHNEOS surveys;][]{2018P&SS..157...82P,
     2019AJ....158..196D, 2019Icar..324...41B}.
  However, compared to the current census of over 20,000 NEAs and more than 860,000
  MB asteroids, these 9000 are only the tip of the iceberg.

  \indent In the current era, these "vermin" have become more appreciated, 
  being serendipitously imaged by large sky surveys from the ultraviolet to far-infrared ranges.
  The most prominent examples of such un-targeted observations
  are the scientific exploitations of the 
  Sloan Digital Sky Survey (SDSS) -- which provides multi-filter photometry 
  in the visible, constraining the composition \citep{2001-AJ-122-Ivezic} --
  and the
  Wide-field Infrared Survey Explorer (WISE) -- which provides mid-infrared photometry and
  hence diameter and albedo determinations \citep{2011ApJ...741...68M}.
  
  \indent However, most imaging surveys do not report moving objects as direct products.
  They must be specifically searched for on released data, generally
  by external teams. The recent works of 
  \citet{2016AA...591A.115P} on the European Southern Observatory (ESO)  Visible and Infrared Survey Telescope for Astronomy (VISTA)
  Hemisphere Survey 
  and \citet{2018AA...610A..21M} on the ESO VST (Very Large Telescope (VLT) Survey Telescope)
  Kilo-Degree Survey
  are typical examples.
  This situation is, fortunately, evolving.
  The current European Space Agency (ESA)
  Gaia mission includes a specific processing dedicated
  to Solar System objects (SSOs) for which a data release already
  occurred \citep{2018-AA-616-Spoto},
  and the upcoming ESA \textsl{Euclid} mission
  and the Legacy Survey of Space and Time (LSST) of the 
  Vera C. Rubin observatory include SSOs
  releases in their respective baselines
  \citep[see][]{2018-AA-609-Carry, 2009-EMP-105-Jones}.

  \indent We focus here on broadband photometry, in the visible, 
  as it provides strong constraints on asteroid composition and classification.
  With over 100,000 reported individual asteroids with five filters in the visible
  \citep[\bandu, \bandg, \bandr, \bandi, and \bandz; ][]{1998AJ....116.3040G},
  the SDSS has been the main source of compositional 
  information for almost two decades.
  Its multi-filter photometry has been analyzed both
  in its own color space
  \citep{2004-MNRAS-348-Szabo, 2005-Icarus-173-Nesvorny}
  and  mapped into a classification compliant with visible
  and near-infrared spectroscopy
  \citep{2010AA...510A..43C, 2013Icar..226..723D}. 
  SDSS photometry has been the source of a variety of studies,
  including
  regarding the selection of candidates for dedicated spectroscopic surveys
  \citep{2008Icar..198...77M, 2014-AA-572-Oszkiewicz,
     2014-Icarus-229-DeMeo,2019-Icarus-DeMeo}, dynamical families and 
  surface aging through space weathering
  \citep{2008-Icarus-198-Parker, 2012-Icarus-219-Thomas,
  2018-Icarus-304-Graves},
  the orbital distribution of hydrated C-types \citep{2012-Icarus-221-Rivkin},
  and the general structure of the asteroid belt \citep{2014Natur.505..629D}.

  \indent Even though the SDSS Moving Object Catalog (MOC)
  has been widely used, its current release, which is called
  the ADR4 \citep{2010PDSS..124.....I} and was published in 2008,
  only contains a fraction of the entire data set.
  First, only \numb{\nummocid} moving sources were identified (i.e., linked with
  a known SSO) out of the \numb{\nummoc} detected moving sources.
  As the number of known SSOs has increased from
  about \numb{\mpcmoc} to \numb{\mpcnow} since
  2008, many of the identifications missing from the ADR4 may be recovered
  (such as was done by \citep{2016Icar..268..340C} who identified \numb{\numsvoid} observations of
  \numb{\numsvoiduniq} asteroids).
  Second, the data selection at the origin of the SDSS MOC imposed
  strict lower and upper limits on the apparent velocity of sources
  between frames. This led to the rejection of any object moving faster than
  0.5 deg/day, that is,   effectively filtering out most NEAs present in the images
  \citep{2014-AN-335-Solano}.
  Finally, the last observing run included in the ADR4 release
  dated from March 2007, while the SDSS imaging survey lasted until November
  2008. 

  \indent The present article aims to increase the number of asteroids
  with SDSS astrometry and multi-filter photometry by identifying previously
  unidentified moving sources and searching for moving sources in 
  frames not included in the latest SDSS MOC release.
  The article is organized as follows.
  In \Autoref{sec:extract} we describe the approach we take
  to identify potential moving objects from the SDSS archive, aiming at
  maximizing completeness.
  We detail the filters applied to the sample
  to increase its purity (i.e., to minimize the number of false-positive
  sources) in \Autoref{sec:clean}.
  The identification of known sources among these
  candidates is described in \Autoref{sec:ident}.
  The completeness and purity of the catalog is
  estimated in \Autoref{sec:purity}.
  We then address suspicious individual measurements
  (astrometry or photometry) in \Autoref{sec:flag}.
  We classify the objects consistently with the 
  \citet{2009Icar..202..160D}
  taxonomic classification in \Autoref{sec:taxo}.
  Finally, we summarize the released asteroid sample in
  \Autoref{sec:conclusion}.

\section{Extraction of Solar System object candidates\label{sec:extract}}

  \indent As noted earlier, the latest release (ADR4) of the SDSS MOC took place
  before the end of operations of the SDSS imaging survey.
  We thus compared the list of \numb{519} runs (number, which identifies the specific scan) \footnote{An SDSS
    run encompasses all the contiguous
    observations along a great circle.}
    included in the ADR4
  with the \numb{832} runs of SDSS Data Release 16 \citep[DR16;][]{2019arXiv191202905A}
  and the \numb{303} runs of Stripe\,82 \citep{2008ApJS..175..297A}, 
  together known as the SDSS Legacy Survey.
  
\begin{figure}[t]
    \centering
    \includegraphics[width=0.99\hsize]{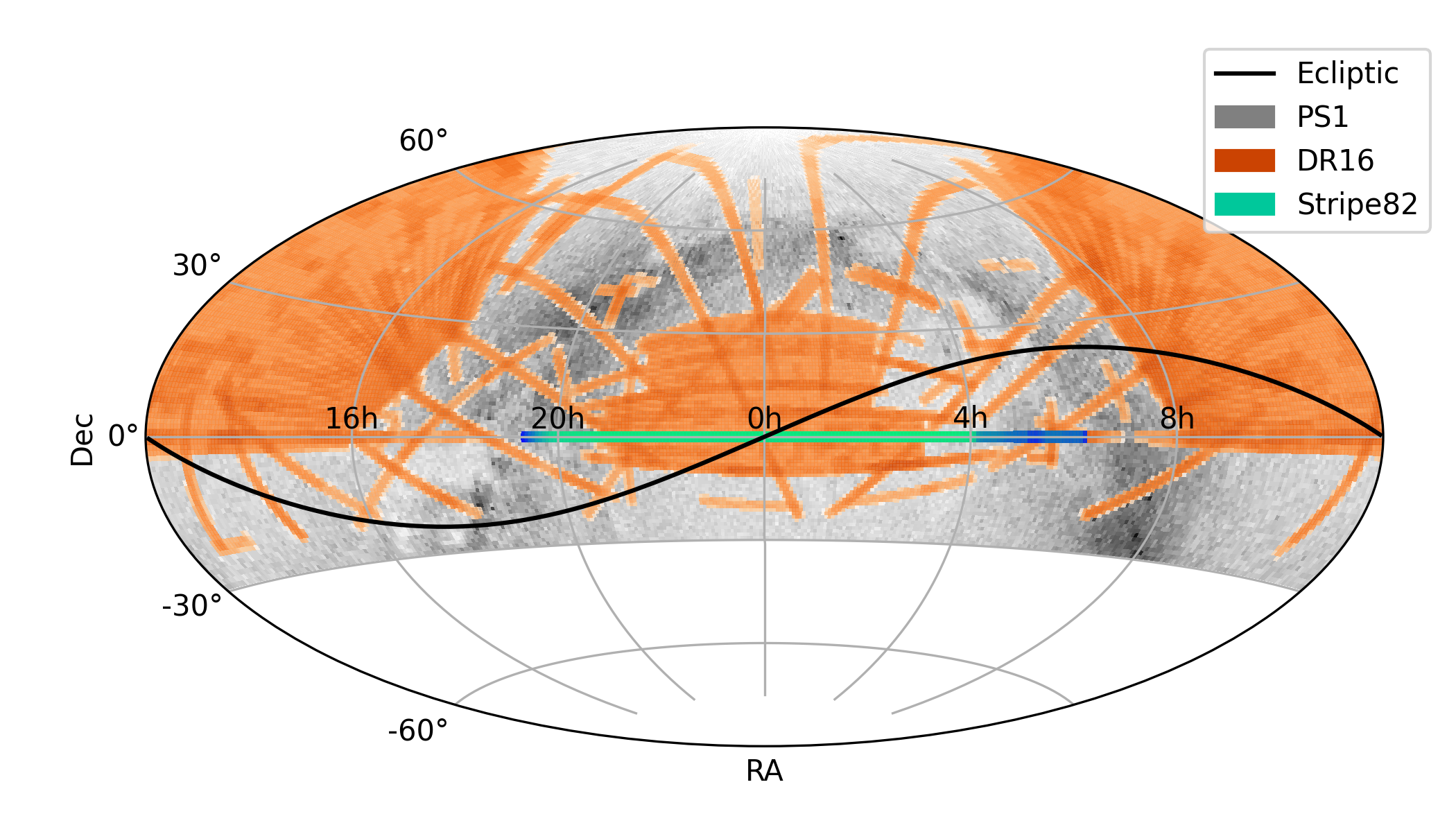}
    \caption{Sky coverage of the main SDSS survey (orange)
      and Stripe\,82 (light blue) in equatorial coordinates (Aitoff projection).
      The background image is generated from the Pan-STARRS stellar catalog.
      \citep{2016arXiv161205560C}.
      The black line represents the ecliptic plane.
      }
    \label{fig:moc}
\end{figure}

  \indent The main SDSS imaging survey covered 7,500\,deg$^2$ in the northern
  Galactic cap and 750 deg$^2$ in the southern Galactic cap, where each
  patch of the sky was imaged in five filters
  in a single visit.
  The 270\,deg$^2$ of Stripe\,82 (a thin long area within
  21h\,$<$\,RA\,$<$\,04h and
  -1.25\degr\,$<$\,DEC\,$<$\,1.25\degr; see \Autoref{fig:moc})
  were repeatedly imaged (about 80 times) as a time-domain survey 
  \citep{2008AJ....135..348S, 2008AJ....135..338F}.
  Both surveys include \numb{80} common runs.
  We only kept \numb{756} runs from DR16, rejecting the runs that were affected by bad weather  (e.g., clouds,
  bad seeing) and thus had a 
  quality score below 0.1.
  In total, we analyzed \numb{938} unique runs, about twice as much as those listed
  in the ADR4.

  \indent The median 5$\sigma$ depth of SDSS photometric observations, based on the formal uncertainties from point spread function (PSF) 
  photometry on point sources, are,
  in AB magnitudes:
  \bandu\,=\,22.15,
  \bandg\,=\,23.13,
  \bandr\,=\,22.20,
  \bandi\,=\,22.20, and
  \bandz\,=\,20.71 \citep{2019arXiv191202905A}.
  We thus set a brightness limit of 22.2 in \bandr 
  \citep[0.7 magnitude fainter than
  the selection criterion of][]{2001-AJ-122-Ivezic}.

  Contrary to the ADR4 extraction
  we did not set an upper limit on the apparent velocity,
  to avoid the rejection of NEAs. 
  Furthermore, as sources moving faster than 0.5\degr/day may appear
  trailed in the images, 
  we did not restrict the object type to stars only, but allowed both 
  stars (\texttt{type=6}) and galaxies (\texttt{type=3}).
  The larger number of runs, combined with our more relaxed selection
  criteria to maximize the completeness
  (the SQL (Structured Query Language)
  code is provided in \Autoref{app:sql} for
  reproducibility).
  leads to a dramatic
  increase in the number of SSO candidates
  compared to the ADR4:
  \numb{\sdssDRfull} in DR16 and \numb{\sdssSTRIPfull} in Stripe\,82.
  
\section{Rejection of false-positive candidates\label{sec:clean}}

  \indent The above extraction returned almost 9 million
  SSO candidates.
  As the selection
  criteria aimed at completeness, many if not most candidates could be stars or
  galaxies (i.e., false-positive identification rather than genuine SSOs).
  We describe below the successive steps we performed to improve the purity of
  the sample by removing these false-positive detections.

  \subsection{Identification of background sources with Pan-STARRS\label{ssec:PS1}}

    The Pan-STARRS catalog provides a homogeneous, multiband survey
    that covers the sky north of declination -30\degr\xspace
    (\autoref{fig:moc}).
    It probes the sky with similar filters, depth, and
    seeing quality as the SDSS \citep{2020ApJS..251....7F}.
    The mean $5\sigma$ point source limiting sensitivities in the Pan-STARRS stacked images reach up to a 23.3 mag limit.
    The epochs of observation separated by years between the SDSS and Pan-STARRS
    provide a practical means to identify fixed-coordinate objects
    (stars and galaxies) among the SSO candidates.
    
    We used the Pan-STARRS Python search API
    (MAST\footnote{\href{http://ps1images.stsci.edu/ps1_dr2_api.html}
    {http://ps1images.stsci.edu/ps1\textunderscore dr2\textunderscore api.html}})
    to compare our catalog of candidates with the Pan-STARRS Data Release 2 (DR2) catalog.
    We excluded candidates matched within 2\arcsec 
    by Pan-STARRS DR2 objects
    if the latter were detected at least twice by Pan-STARRS.
    After this filtering, the number of 
    SSO candidates dropped to \numb{\ssops}.

 \subsection{Exclusion of duplicates\label{ssec:duplicate}}

    \indent The SDSS \texttt{PhotoObjAll} catalog we queried
    (\Autoref{app:sql})
    contains duplicated measurements in some cases (i.e., the
    same observation could appear a few times in the catalog). 
    This is a consequence of several factors. 
    
    The first is the methodology of SDSS photometry.
    Identical objects could be measured in different fields because part of the frames has cross-covering. The second is the fact that while the Stripe82 and DR16 catalogs have the same runs, 
    they  were processed by different versions of the SDSS photometry pipeline and 
    therefore could have slightly different photometry.
    The third factor is due to the moving nature of observed SSOs.
    For a high apparent rate, the same object can be identified
    as two or more stationary sources
    (see \Autoref{ssec:fast} for a description of our extraction of photometry for
    fast-moving asteroids such as NEAs).
    For each duplicated entry, we grouped objects by run number within
    a 5\arcsec radius and selected the most accurate
    photometry for each band.

  \subsection{Identification of high-proper-motion stars with \textsl{Gaia}\label{ssec:gaia}}

    \indent While the cross-match with Pan-STARRS allows the identification and
    rejection of most stationary sources, high-proper-motion stars may have
    moved farther than our cross-match radius of 2\arcsec in 20 years.
    We selected all stars listed in \textsl{Gaia} DR2
    \citep{2018AA...616A...1G} with proper motion above 0.05 arcsec/year, 
    fainter than G=14.0, and with declination above -30\degr.
    For each of the \numb{2,278,339} stars we retrieved, we computed its position at 
    the SDSS mid-observing epoch, January 1, 2004 
    (see \Autoref{app:sql}).
    We then cross-matched this catalog of high-proper-motion stars with the 
    catalog of SSO candidates.
    We identified \numb{67} additional stars missed by the Pan-STARRS comparison
    above and rejected them from the sample.

  \subsection{Rejection of dubious frames\label{ssec:frames}}

    \indent Among all SDSS fields, we found frames with a suspiciously high number of SSOs.
    A visual inspection of these fields revealed issues with the astrometry
    across the different filters. As a result, sources have
    different positions in each filter, leading to erroneous
    classification as moving sources by the SDSS software.
    We removed \numb{312} such fields from DR16 and \numb{414}  from Stripe82.
    We also removed all \numb{300} frames of Run 2505, which were totally crowded
    by Milky Way stars.
    In this step, we rejected \numb{6701} candidates from our catalog.

\section{Identification of known Solar System objects\label{sec:ident}}

  \indent The extraction and selection presented in the previous section
  resulted in a catalog of \numb{\ssototprev} observations.
  To identify the detected objects, we performed \numb{\sdssDRFields}
  (DR16) and \numb{\sdssStripeUniqueFields} (Stripe 82)
  cone-search 
  queries with \skybot \citep[the Sky Body Tracker;][]{2006-ASPC-351-Berthier}, one for each 
  of the frames (13.5$^\prime$\,$\times$\,9.0\arcmin) included in our catalog.
  \skybot is a Virtual Observatory web service
that provides cone-search
utilities for SSOs. It is thus a
useful tool for seeking and identifying moving objects in astronomical images.

\begin{figure}[t]
    \centering
    \includegraphics[width=0.99\hsize]{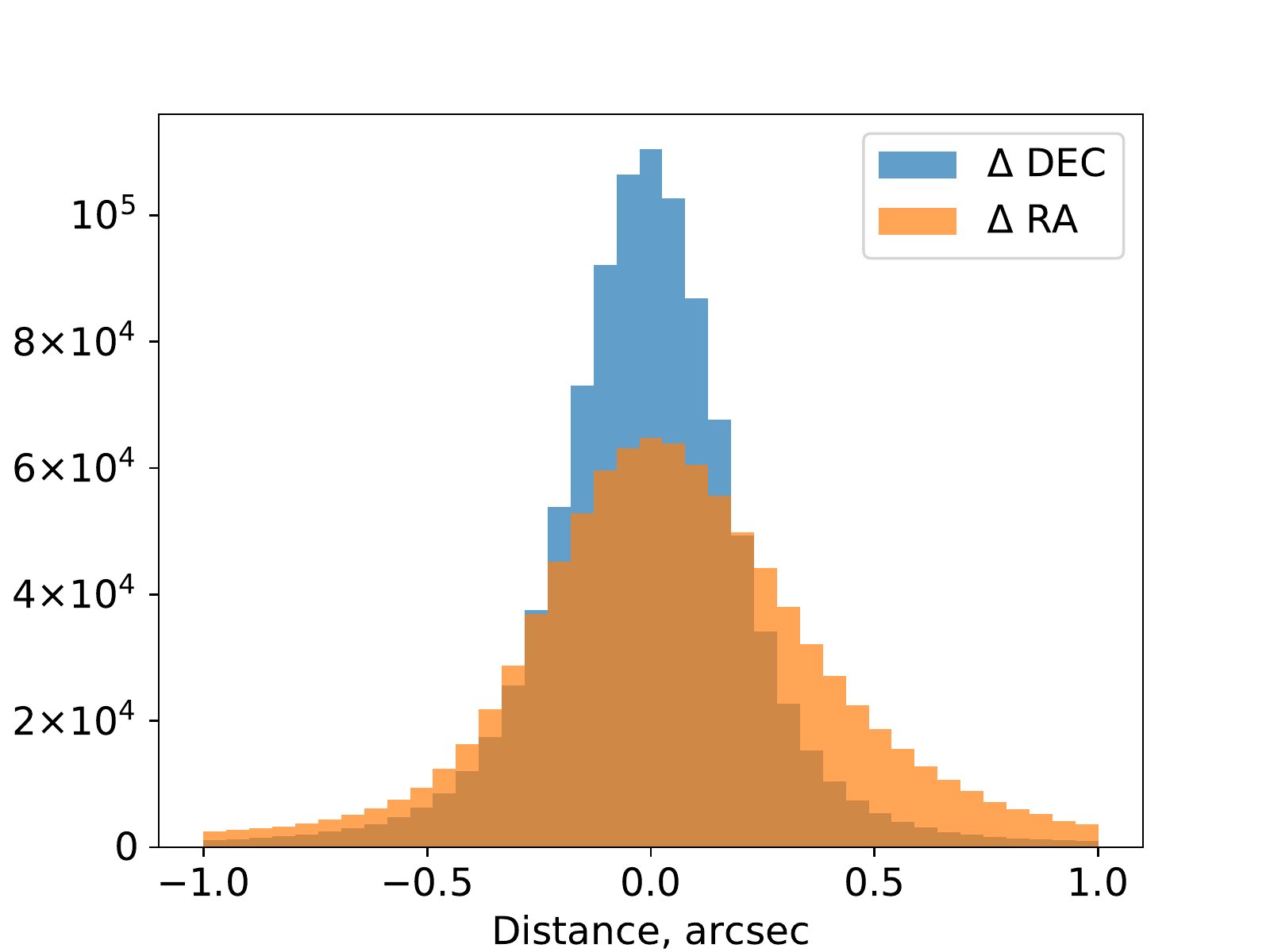}
    \caption{Distribution of distances between \skybot predictions
    and SDSS measurements.
    }
    \label{fig:ang_dist}
\end{figure}

\begin{figure}[t]
    \centering
    \includegraphics[width=0.99\hsize]{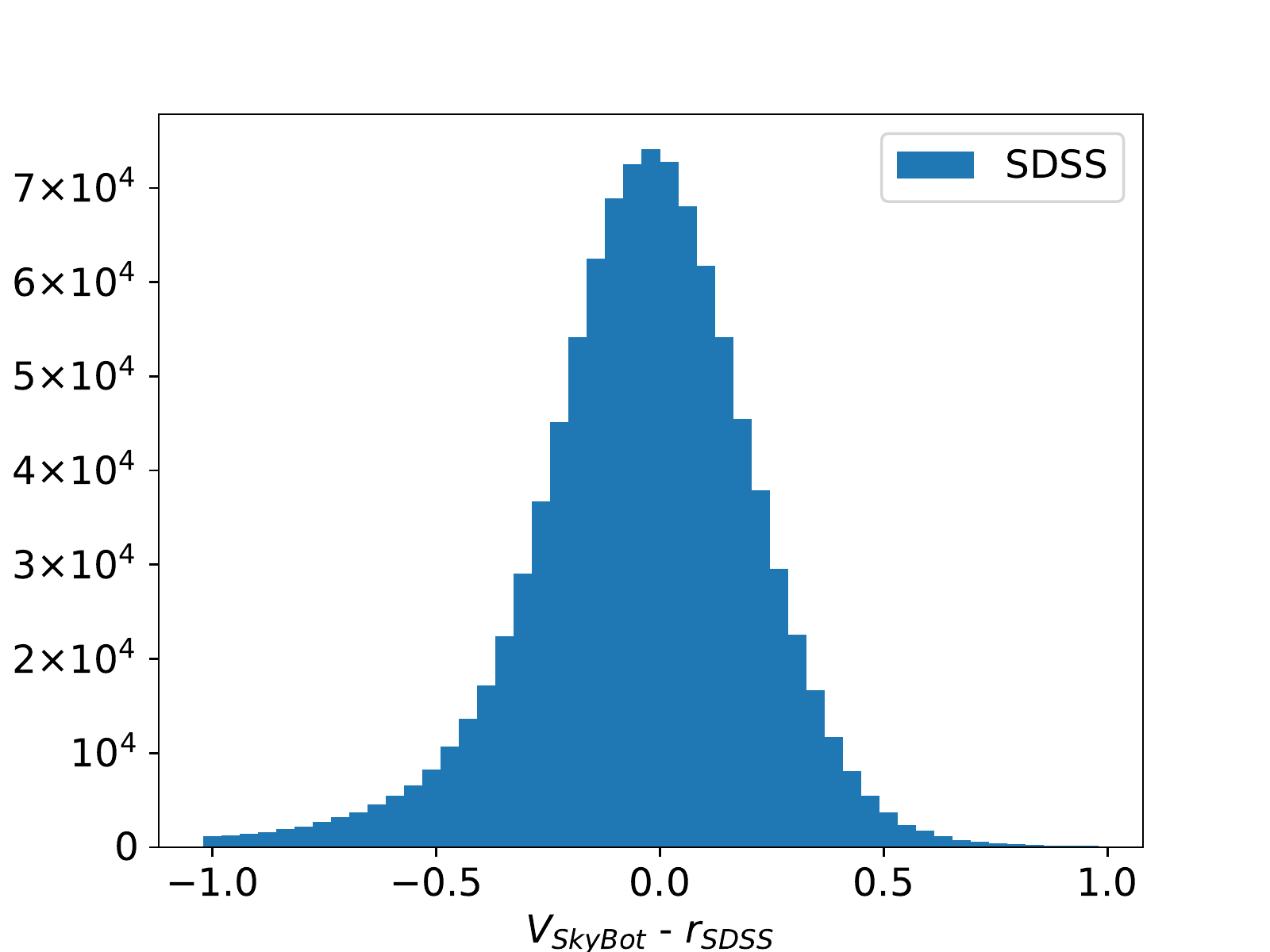}
    \caption{Distributions of \skybot V magnitudes and transformed SDSS SSOs.
    }
    \label{fig:Vmag}
\end{figure}

  \indent In \numb{\skytot} of our cases, 
  a known SSO was predicted by \skybot to be located near
  the source we extracted from the SDSS archive (\Autoref{sec:extract}).
  The mean distance between the prediction and the detection is 
  \numb{-0.05\,$\pm$\,0.45}\arcsec in right ascension and 
  \numb{-0.01\,$\pm$\,0.22}\arcsec in declination 
  (\Autoref{fig:ang_dist}).
  The V-\bandr color, presented in 
  \Autoref{fig:Vmag}, is \numb{0.01\,$\pm$\,0.24}
  \citep[i.e., slightly redder than Solar color;][]{2018ApJS..236...47W}. 
  We associated the \skybot prediction with the SDSS measurement
  whenever the angular distance was smaller than
  \numb{30\arcsec}, 
  the $|V-\bandr|$ magnitude difference was smaller than
  \numb{5}, and
  the difference in apparent motion was within
  \numb{5\arcsec/h}.
  
  \indent In \numb{\ssoNotInMoving} other cases, a known SSO was predicted by
  \skybot but did not correspond to any of the 
  \numb{\ssototprev} sources we extracted (\Autoref{sec:extract}).
  For each of these predictions,
  we extracted the closest source in the SDSS archive.
  The majority of these predictions
  have small apparent rates (hence were not marked as moving objects by the SDSS pipeline).
  We then applied the suite of filters described above (\Autoref{sec:clean}) and added all these recovered sources to the catalog.

\begin{figure}[t]
    \centering
    \includegraphics[width=0.99\hsize]{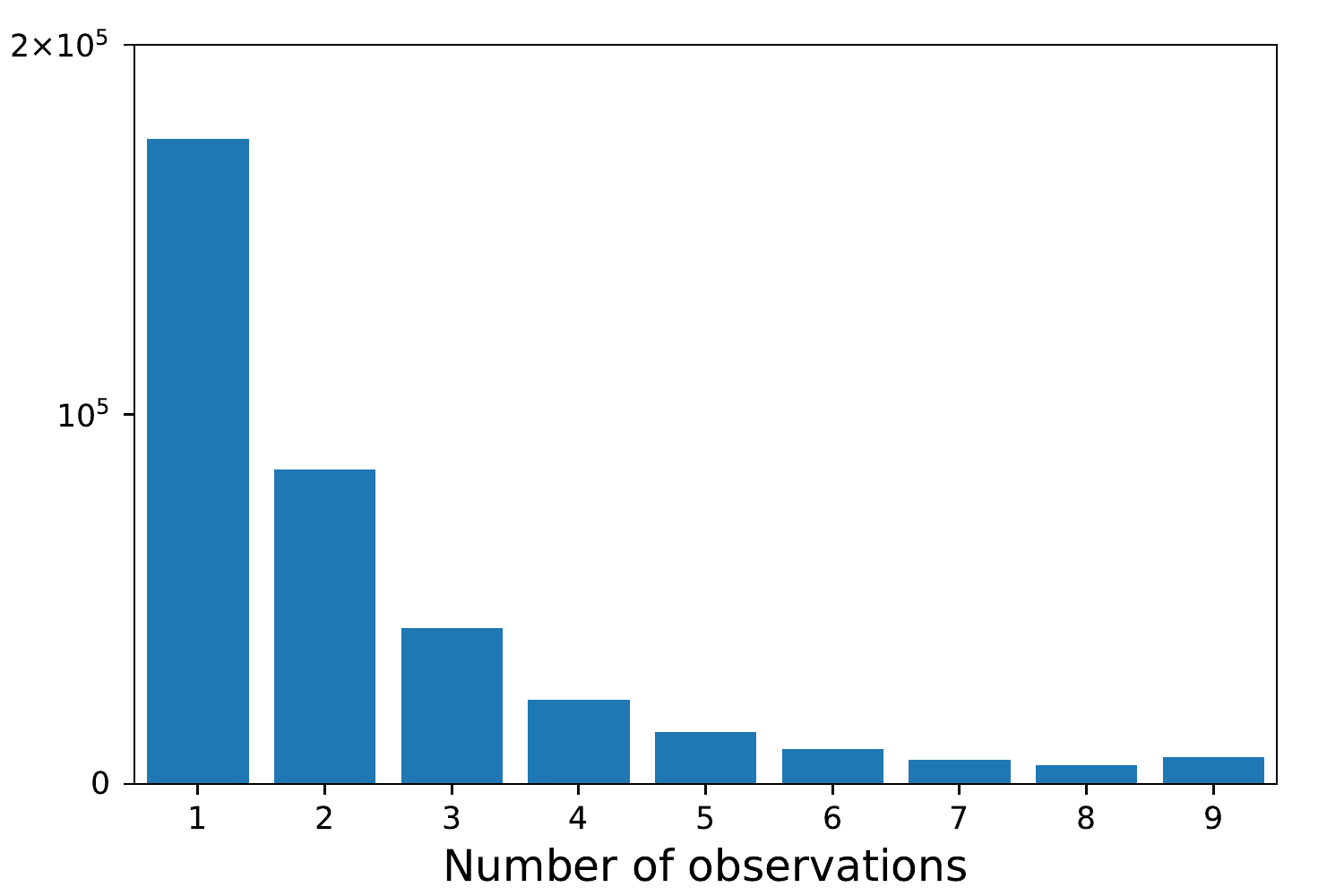}
    \caption{Distribution of the number of observations per SSO.
    }
    \label{fig:ast_observation}
\end{figure}

  \indent In total, we identified
  \numb{\ssoknown} observations of \numb{\ssouniq} known SSOs. 
  The distribution of the number of asteroid observations is presented
  in \Autoref{fig:ast_observation}. 
  Nearly half of the asteroids were observed only once; 
  however, the mean value of the observations of individual asteroids is 2.7. 
  Moreover, \numb{13,882} asteroids have more than ten observations.
  
  The catalog contains \numb{\ssotot} observations, including \numb{\ssounknown} observations of moving 
  objects not linked with any known SSO.
  Most of the identified SSOs are located in the main asteroid belt. 
  All dynamical classes are, however, present in the catalog, 
  including comets (\Autoref{tab:dynclass}).

\begin{table}[ht]
    \centering
    \caption{Number of observations ($\#_{\textrm{obs}}$) and
    unique SSOs ($\#_{\textrm{SSOs}}$) for each dynamical class.
    }
    \label{tab:dynclass}
    \begin{tabular}{llrr}
        \hline
        \hline
        \multicolumn{2}{c}{Dynamical class} &
          $\#_{\textrm{SSOs}}$ & $\#_{\textrm{obs}}$ \\
        \hline
        NEA & Aten   & 52 & 68 \\
        NEA & Apollo & 836 & 1451 \\
        NEA & Amor   & 764 & 1,355 \\
     \noalign{\smallskip}
        \multicolumn{2}{l}{Mars-Crosser} & 4,242 & 9,024 \\
        \multicolumn{2}{l}{Hungaria} & 6,362 & 12,841 \\
    \noalign{\smallskip}
        MB & Inner  & 105,273 & 276,899 \\
        MB & Middle & 134,878 & 370,154 \\
        MB & Outer  & 119,392 & 337,668 \\
        MB & Cybele & 1,722 & 5,070 \\
        MB & Hilda  & 1,923 & 5,021 \\
     \noalign{\smallskip}
        \multicolumn{2}{l}{Trojan} & 3,929 & 8,721 \\
        \multicolumn{2}{l}{Centaur} & 123 & 522 \\
     \noalign{\smallskip}
        \multicolumn{2}{l}{KBO} & 1,024 & 3,143 \\
        \multicolumn{2}{l}{Comet} & 233 & 420 \\
        \hline
        \multicolumn{2}{l}{Total} & \ssouniq & \ssoknown\\
        \hline
    \end{tabular}
\end{table}

\section{Purity and completeness\label{sec:purity}}

  \indent We estimated the completeness of our catalog using the 
  association with known SSOs presented above (\Autoref{sec:ident}).
  The SSOs predicted by \skybot but not present in our catalog
  provide an estimate of its completeness, albeit not an unbiased one.
  The current census of SSOs may itself be biased
  \citep{2018AA...610A..21M}, and 
  a 100\% completeness with known SSOs may not necessary imply a full completeness
  of sources on sky.
  As an illustration, about a third of all SSOs reported in the present catalog
  are not identified.
  Hence, this comparison provides an estimate on the fraction of objects removed
  by the different filters we applied in \Autoref{sec:clean}.

\begin{figure*}[t]
    \centering
    \includegraphics[width=0.8\hsize]{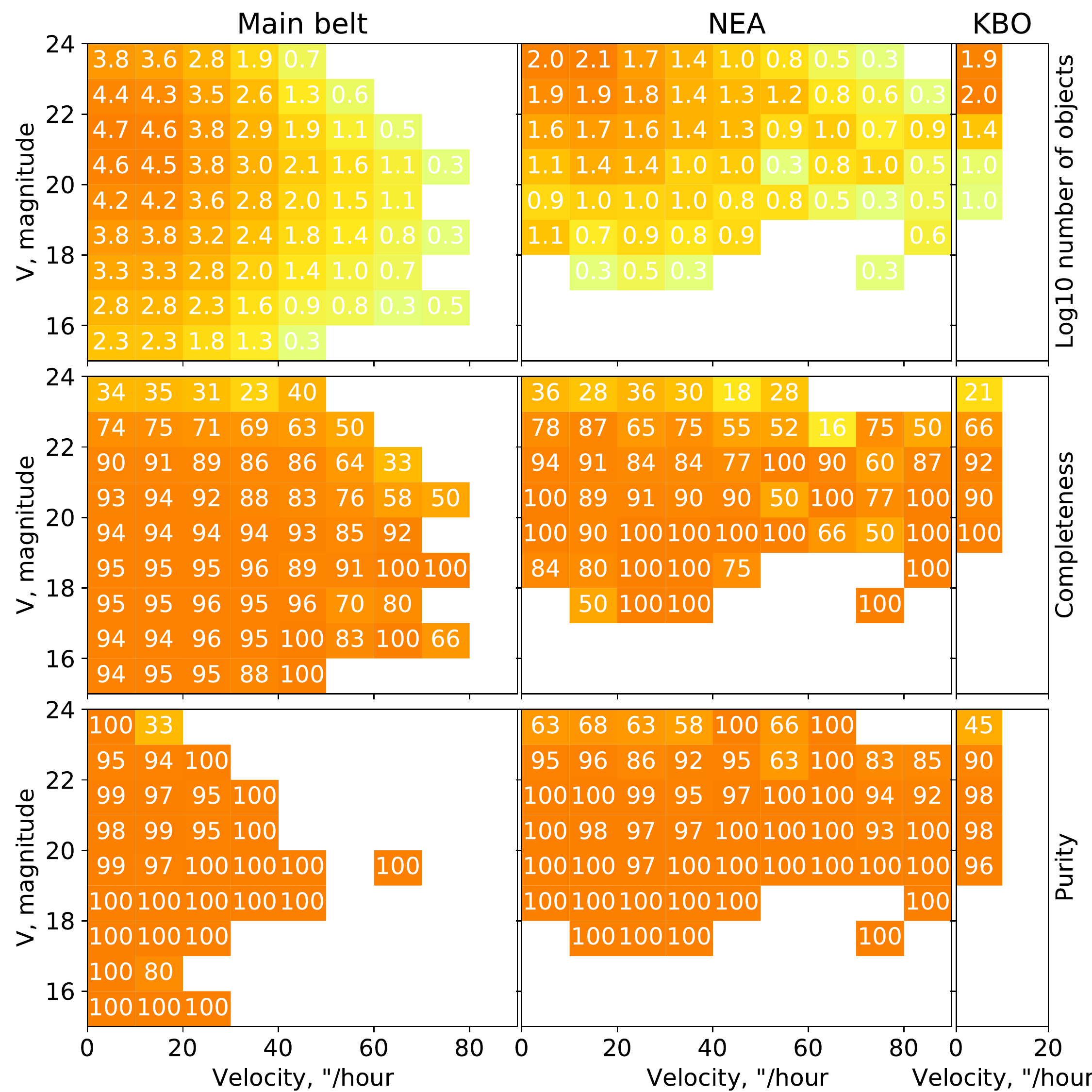}
    \caption{Distribution of the population (top), 
    completeness (middle), and purity (bottom)
    of asteroids (MB asteroids and NEAs)
    and KBOs.
    }
    \label{fig:completeness}
\end{figure*}

  \indent We treated  the cases of MB asteroids, NEAs, 
  and Kuiper belt objects (KBOs) separately owing to their vastly different apparent motion.
  In all cases, we studied the completeness as a function of the apparent magnitude
  (as reported in V by \skybot) and motion (in \arcsec/h).
  We present the estimated completeness in \Autoref{fig:completeness}.
  The completeness is typically around 95\% until V$\approx$22, where it drops to about 70\% at
  V$\approx$23, and then to 30\% at V$\approx$24.
  Such a completeness is similar to that of the ADR4 
  \citep{2001-AJ-122-Ivezic}.
  
  \indent We estimated the purity of the catalog by visually inspecting RGB (red, green, blue)
  images made by combining the \bandg, \bandr, and \bandi frames (similar to the insets of
  \Autoref{fig:linear}). 
  We inspected all KBOs, 
  \numb{2000} randomly selected MB asteroids,
  and \numb{2915} NEAs.
  The purity of the catalog is typically above 97\% down to magnitude V$\approx$23,
  where it suddenly drops (\Autoref{fig:completeness}).
  This represents a clear improvement with respect to the ADR4, for which
  a 94\% purity was quoted for MB asteroids, dropping to 
  50\% for both fast and slow motion SSOs \citep{2001-AJ-122-Ivezic}.

\section{Identification of spurious measurements\label{sec:flag}}

  The filters described in \Autoref{sec:clean} 
  limit the number of false positives 
  among the \numb{\ssotot}
  observations we report (\Autoref{sec:purity}).
  We describe below further tests for identifying potential issues with
  the astrometry or photometry of measurements in individual bands.
  
  \subsection{Spurious astrometry from linearity of motion\label{ssec:motion}}

\begin{figure*}[t]
    \centering
    \includegraphics[width=0.99\hsize]{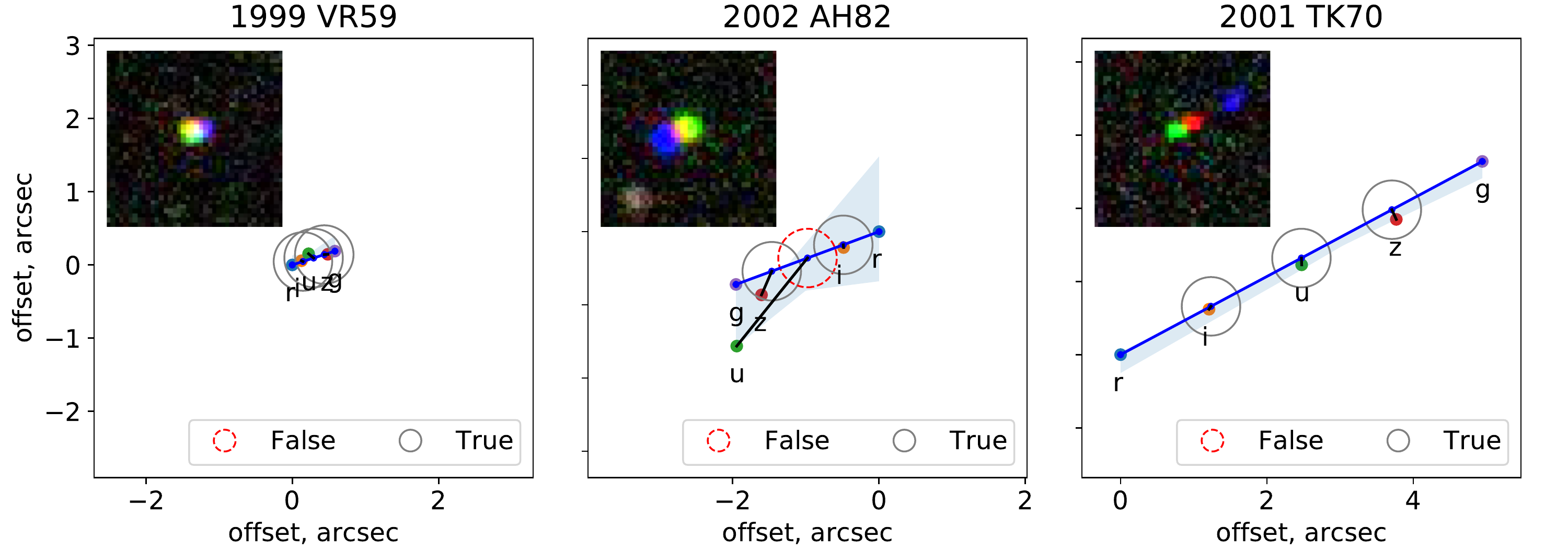}
    \caption{Examples of asteroid signatures in SDSS frames:
    slow-moving (0.05\,\degr/day, \textsl{left}),
    typical motion ($\sim$0.25\,\degr/day, \textsl{center}),
    and fast-moving (0.5\,\degr/day, \textsl{right}). 
    For each, the observed position offsets (red points) and expected position offsets (blue points)
    with respect to \bandr are
    shown.
    The solid gray and dashed red circles with 0.4 arcsec (1 pixel) radii
    indicate validated and spurious astrometry, respectively.
    For each object, we also present a \bandg\bandr\bandi tri-color 
    image at the top-left corner. 
    }
    \label{fig:linear}
\end{figure*}

    \indent As the SDSS imaging camera scanned the sky, the same 
    region of the sky was sequentially imaged in
    \bandr, \bandu, \bandi, \bandz, and \bandg, with a time interval of
    17.7\,s between each 54\,s exposure frame. 
    From the reported position offsets in each band with respect to the position
    in \bandr, we built the footsteps of each SSO to check 
    the linearity of its motion \citep{2019AC....2800289M}.
    We used the Siegel estimator, a linear regression that is robust to outliers     \citep{medianregress}.
    We find that \numb{92}\% of the SSOs have a coefficient of determination,
    $R^2$, larger than 0.9.
    Considering that an offset position in a single band significantly
    lowers $R^2$, such high values provide a strong indication of linear motion,
    and hence a confirmation of the genuineness of these SSOs.
    We used this criterion to identify spurious astrometry in the
    catalog (see \Autoref{app:cat}).
    Using \bandg and \bandr as reference (the two deepest filters;   \Autoref{sec:extract}), we predicted the expected position in 
    \bandz, \bandu, and \bandi.
    We marked as dubious each position in these bands that is located farther
    than one pixel (0.4\arcsec) from the prediction (\Autoref{fig:linear}).

  \subsection{Identification of slow-moving SSOs\label{ssec:slow}}
  
    \indent The extraction of SSO candidates (\Autoref{sec:extract})
    imposes a minimum apparent motion of 0.05\degr/day to reject
    all stationary sources, such as stars and galaxies
    \citep[similar to the ADR4 extraction;][]{2001-AJ-122-Ivezic}. 
    This threshold converts to a minimum change in position 
    of 0.6\arcsec between
    the \bandr and \bandg frames, which provide the largest
    temporal baseline, 286.8\,seconds.
    This effectively precludes the identification of most moving
    objects from the outer Solar System
    \citep[only 41 SSOs with a semimajor axis above 5.5 au were reported
    by][]{2001-AJ-122-Ivezic}.
    
\begin{figure}[!]
      \centering
      \includegraphics[width=0.99\hsize]{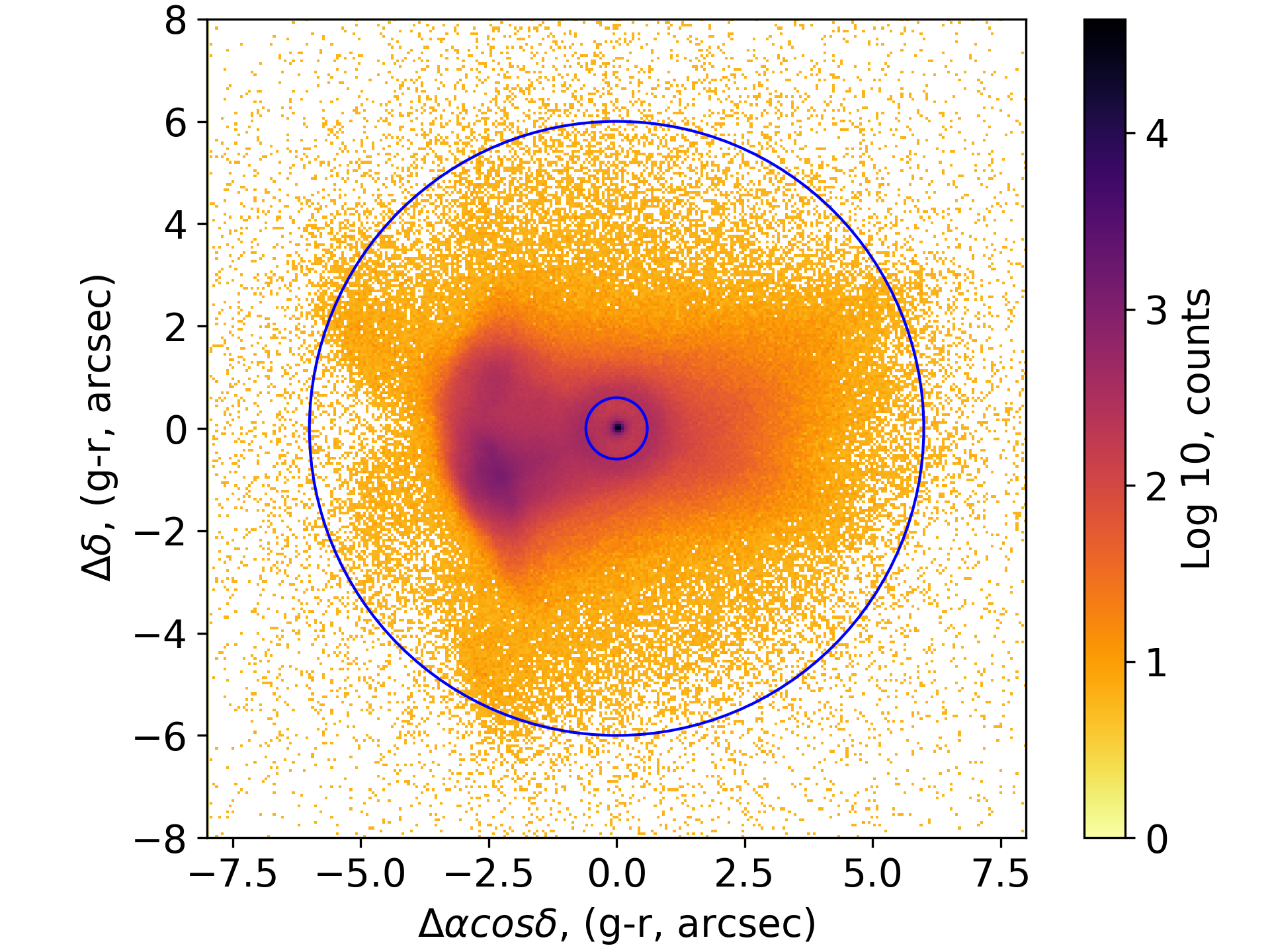}
      \caption{Position offset between \bandr and \bandg.
        The inner black circle represents the lower limit of 0.05\degr/day that we
        use to extract SSO candidates.
        The outer blue circle shows the upper limit of
        0.5\degr/day from the ADR4 that we do not use here.}
      \label{fig:offset}
    \end{figure}

    \indent As is visible in \Autoref{fig:offset}, however,
    \numb{\ssoslow} objects do not display apparent motion.
    These can be slow-moving objects, such as KBOs 
    or asteroids at their turnaround point between direct and 
    retrograde motion. 
    However, the false-positive rate may be higher among these objects
    because the identification of these sources as moving objects 
    is more prone to uncertainty (the linearity of motion cannot be checked, for instance).
       
    We thus marked \numb{\ssoslowbad} objects in the catalog
    (\Autoref{app:cat}) as suspicious:
    Their change in position between \bandg and \bandr is less than 0.6\arcsec and
    there are no known SSOs predicted within 5\arcsec.
    We, however, did not flag out the 
    \numb{\knownslow} slow-moving SSOs associated with a \skybot counterpart
    (\Autoref{sec:ident}).

  \subsection{Retrieving the photometry of fast-moving SSOs\label{ssec:fast}}

    \indent As a feature of the SDSS catalog creation, moving objects 
    with a large apparent motion may not be recognized as such
    \citep{2001-AJ-122-Ivezic}.
    Their successive appearances in the different filters may be cataloged
    as different SDSS sources.
    We thus applied an improved source extraction for \numb{\fastnumber} SSOs 
    with a proper motion above 60\arcsec/h.
    This apparent velocity corresponds to an apparent displacement of
    1\arcsec (the mean seeing value) over the 54\,s integration of each
    filter.
    
    The extraction procedure includes extracting all sources
    in the predicted moving distance 
    from the SDSS catalog for each band. 
    We then checked sources along with predicted positions
    and combined them into one moving source.    
    
    For each filter, we replaced the astrometry and photometry from the previous
    extraction with those of the SDSS source located the closest to the
    \skybot prediction. We then reapplied all the filters described in
    \Autoref{sec:clean}.

  \subsection{Spurious photometric uncertainty\label{ssec:unc}}

\begin{figure}[t]
    \centering
    \includegraphics[width=0.99\hsize]{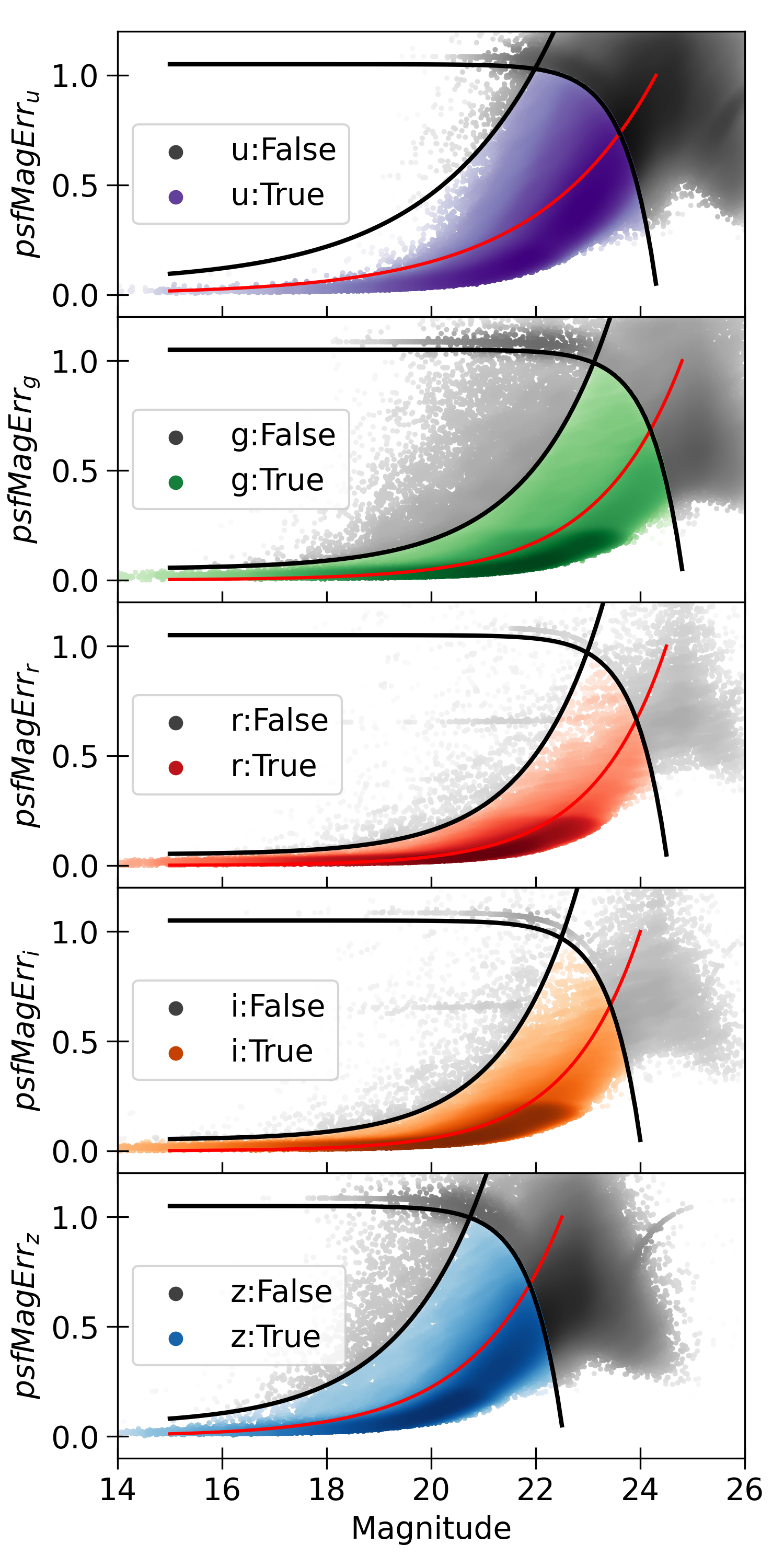}
    \caption{Photometric uncertainty of SSOs in all bands.
    The photometry of the gray points is marked as suspicious.
    The black curves represent the acceptance limits and the red curve
    the global trend (see text). 
    }
    \label{fig:mag_err}
\end{figure}

    \indent As is visible in \Autoref{fig:mag_err}, the distribution of photometric
    uncertainties in each filter is too complex to be due to photon noise only.
    There are in particular two spurious behaviors, which are present in all filters.
    We marked the photometry as suspicious in the catalog (\Autoref{app:cat}) using
    the following empirical threshold:
    \begin{equation}
      \textrm{psfMagErr}_{\textrm{max}} \overset{\text{min}} = \begin{cases}
             1.05 - \textrm{Exp}\left[\frac{\textrm{psfMag} - \textrm{Mag}_\textrm{lim}}{0.6} \right]\\
             \textrm{exp}\left[\frac{\textrm{psfMag} - (\textrm{Mag}_\textrm{lim}-\sigma)}{\sigma}\right]+0.05
         \end{cases}
      \label{eq:photunc}
    ,\end{equation}

    \noindent where the limiting magnitude, $\textrm{mag}_{\textrm{lim}}$, and the
    curvature, $\sigma$, are filter dependent and listed in \Autoref{tab:mag_constrains}.
   
    The first criterion in \Autoref{eq:photunc} identifies the 
    overemphasized uncertainty value of 
    $\sim$1.1 for bright sources, which exponentially decreases to 0 for faint
    sources. 
    The second marks the photometry with 
    uncertainties too large to be meaningful, including 
    the overrepresented uncertainty value of 0.6 mag.
    We set the $\sigma$ curvature by
    fitting the following equation to each band (red curves in 
    \Autoref{fig:mag_err}):
    \begin{equation}
      \textrm{psfMagErr} = \exp{\frac{\textrm{psfMag}-\textrm{mag}_{\textrm{lim}}}{\sigma}}
    .\end{equation}As expected, the largest number of observations with suspicious 
    photometry is presented in the \bandu and \bandz bands, while
    almost all observations in the \bandr and \bandi bands have accurate photometry
    (\Autoref{tab:mag_constrains}).

\begin{table}[t]
    \centering
    \caption{Parameters ($\textrm{mag}_{\textrm{lim}}$, $\sigma$)
      for the selection of photometric uncertainties. We also report the
      fraction of measurements marked as suspicious.}
      \label{tab:mag_constrains}    
    \begin{tabular}{lccccc}
        \hline
        \hline
         Parameter & \bandu & \bandg & \bandr & \bandi & \bandz \\
         \hline
         $\textrm{mag}_{\textrm{lim}}$ & 24.3 & 24.8 & 24.5 & 24.0 & 22.5\\
         $\sigma$ & 2.27 & 1.59 & 1.40 & 1.40 & 1.67 \\
         Suspicious & 46\% & 14\% & 2\% & 2\% & 31\% \\
         \hline
    \end{tabular}
\end{table}
 
  \subsection{Identification of spurious colors\label{ssec:color}}

\begin{figure}[ht]
    \centering
    \includegraphics[width=0.99\hsize]{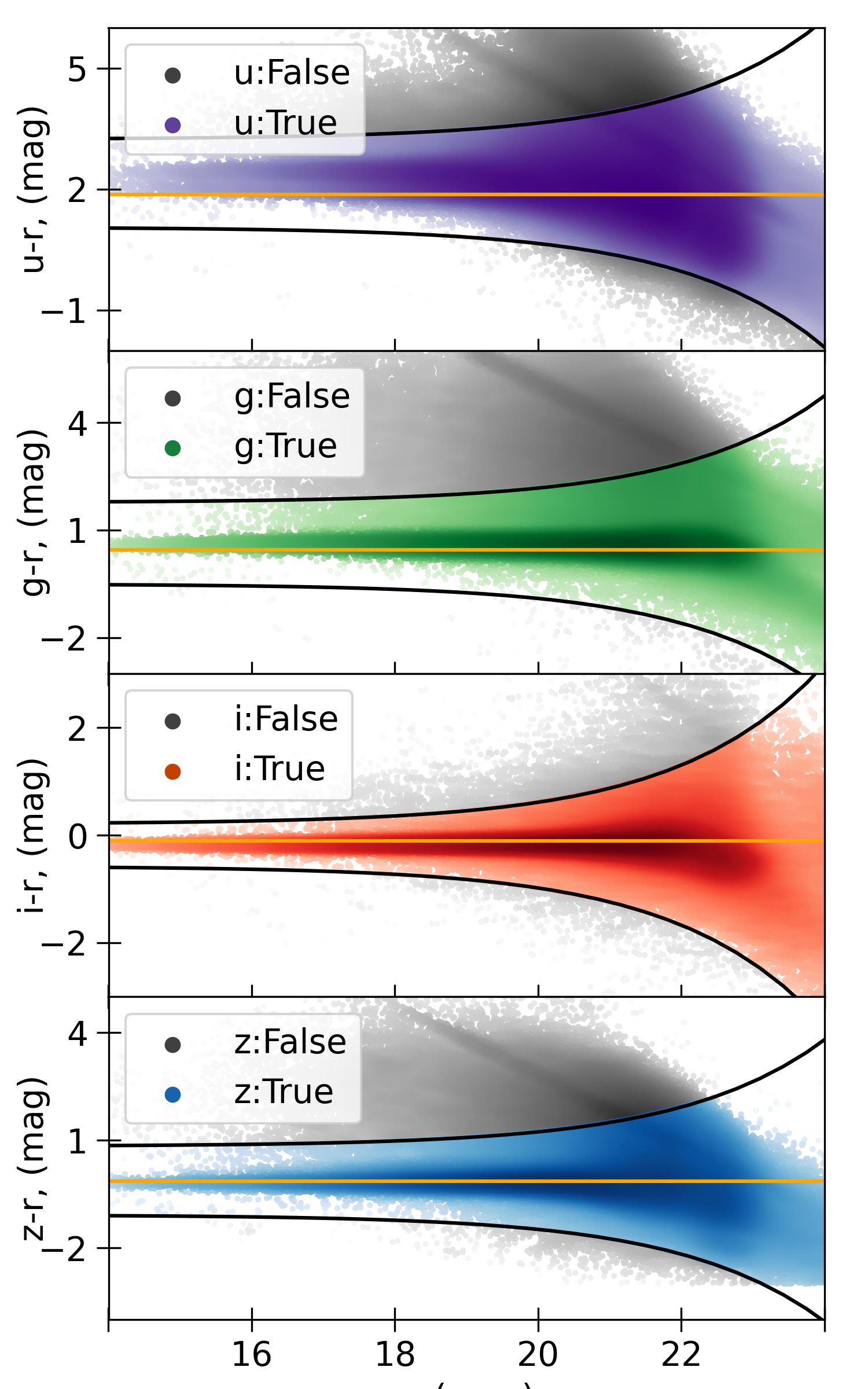}
    \caption{Distribution of
      \bandu-\bandr, \bandg-\bandr, \bandi-\bandr, and
      \bandz-\bandr colors as a function of magnitude.
      The rejected measurements are displayed
      in gray. An orange horizontal line demonstrates the Sun color.}
    \label{fig:color_cond}
\end{figure}

    \indent We performed a final test on the reported photometry.
    In a first approximation, SSOs display Solar colors, and
    a significant departure from these colors can be
    used to identify potential issues with the photometry not 
    identified in the previous steps.
    Using \bandr as a reference, we analyzed how colors vary with 
    magnitude (\autoref{fig:color_cond}). 
    The spread of colors increases with magnitude due to increasing
    photometric uncertainty.
    We flagged photometry resulting in a color
    farther than 3\,$\sigma$ from the median color
    (\autoref{app:cat}).
    The fraction of measurements marked as such amounts to
    7\%, 11\%, 1\%, and 8\% for \bandu, \bandg, \bandi, and \bandz,
    respectively. 

\section{Taxonomic classification\label{sec:taxo}}

  \indent We used the photometry to classify the asteroids within a scheme
  consistent with the widely used Bus-DeMeo taxonomy
  \citep{2009Icar..202..160D}. 
  We followed the approach in \citet{2013Icar..226..723D}, according to which strict boundaries
  in the overall spectral slope and depth of the one-micron band are used to
  classify the SSOs into ten broad complexes:
  A, B, C, D, K, L, Q, S, V, and X.
  We, however, introduced several improvements.
  
\begin{figure*}[t]
    \centering
    \includegraphics[width=0.95\hsize]{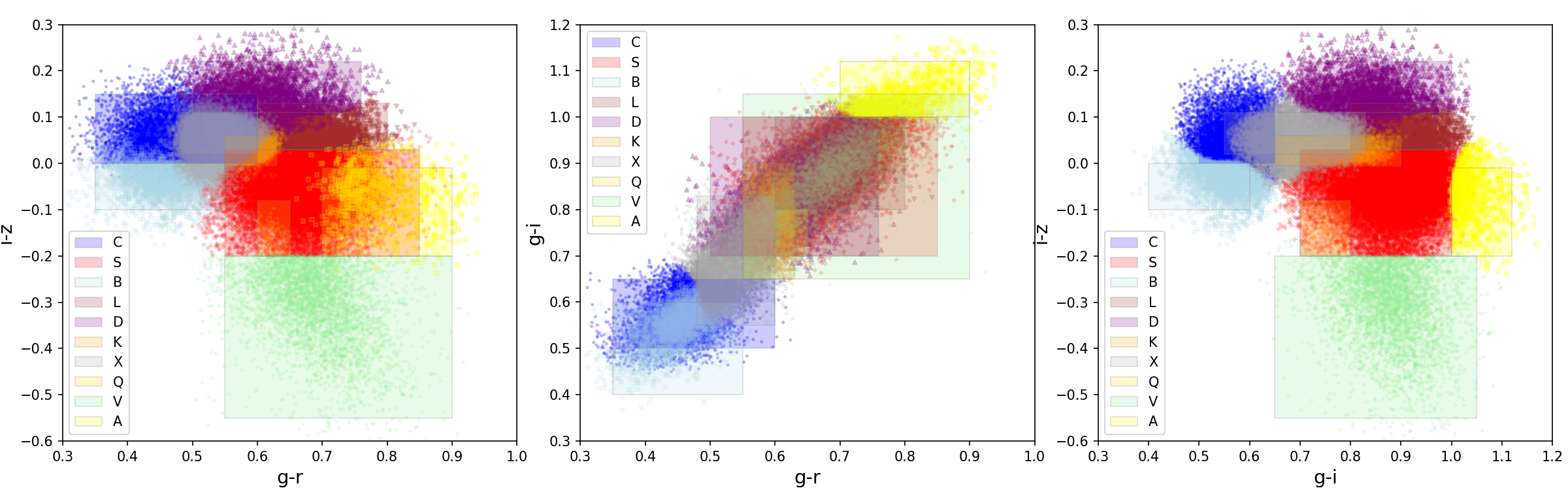}
    \caption{Taxonomy of SSOs.
    Boxes are boundaries of the taxonomic classes. 
    Color points mark the classification of the 72,043 asteroid measurements that have color uncertainties smaller than 0.05 mag.}
    \label{fig:taxo_bound}
\end{figure*}

  \paragraph{Magnitude versus reflectance.} 
  We used the SDSS magnitudes, similarly to what was done in \citet{2010AA...510A..43C},
  rather than converting them
  into spectral reflectance, as was done by \citet{2013Icar..226..723D}.
  Our aim was to remain as close as possible to the original data.
  Moreover, this conversion requires the colors of the
  Sun in the SDSS bands, the estimates of which differ between 
  authors (see, e.g., 
  the SDSS estimates\footnote{\url{https://www.sdss.org/dr16/algorithms/ugrizvegasun/}},
  \citealt{2006MNRAS.367..449H}, or
  \citealt{2012ApJ...752....5R}).
  We thus converted the class boundaries from
  \citet[][Table 3]{2013Icar..226..723D}
  into limits in the 3D \bandg-\bandr, \bandr-\bandi, and \bandi-\bandz 
  color space (\Autoref{tab:taxo_bound}, \Autoref{fig:taxo_bound}).
  As in \citet{2013Icar..226..723D}, some classes overlap in the 3D
  color space. 
  We, however, deal differently with these overlaps, as described below.

\begin{table}[]
    \centering
    \caption{Complex color boundaries.}
    \label{tab:taxo_bound}    
    \begin{tabular}{ccccccc}
    \hline
    \hline
    complex & $gr_{min}$ & $gr_{max}$ & $iz_{min}$ & $iz_{max}$ & $gi_{min}$ & $gi_{max}$ \\
    \hline
    A & 0.70 & 0.90 & -0.20 & 0.03 & 1.00 & 1.10 \\
    B & 0.30 & 0.55 & -0.10 & 0.00 & 0.20 & 0.60 \\
    C & 0.35 & 0.60 &  0.00 & 0.15 & 0.50 & 0.65 \\
    D & 0.53 & 0.85 &  0.11 & 0.34 & 0.70 & 1.00 \\
    K & 0.55 & 0.63 & -0.00 & 0.06 & 0.70 & 0.95 \\
    L & 0.60 & 0.85 &  0.06 & 0.13 & 0.80 & 1.00 \\
    Q & 0.60 & 0.65 & -0.20 & 0.08 & 0.60 & 0.85 \\
    S & 0.55 & 0.85 & -0.20 & 0.06 & 0.65 & 1.00 \\
    V & 0.55 & 0.90 & -0.70 & 0.20 & 0.65 & 1.00 \\
    X & 0.48 & 0.60 &  0.04 & 0.11 & 0.55 & 0.85 \\
    \hline
    \end{tabular}
\end{table}

  \paragraph{Accounting for uncertainties.}
  For each observation, we computed the volume it occupies in the 
  3D color space (\bandg-\bandr, \bandg-\bandi, \bandi-\bandz) based
  on a 3D Gaussian distribution, whose $\sigma$ are set to color uncertainties.

  We then computed a score for each class, $k$, based on 
  the volume of the intersection between each observation 3D Gaussian
  with the space occupied by each taxonomic complex (\Autoref{fig:taxo_bound}), 
  normalized by the Gaussian volume:
  
  \begin{equation}
    V_{\sigma} = \prod_{j=1}^3{
    \left(erf\left[\frac{b_j-\mu_j}{\sqrt{2}\sigma_j}\right] - erf\left[\frac{a_j-\mu_j}{\sqrt{2}\sigma_j}\right] \right)}
  ,\end{equation}
  
  \noindent where
  $erf(z)$ is the error function,  $erf(z) = \frac{2}{\sqrt{\pi}}\int_{0}^{z}{e^{-t^2}dt }$, 
  the index $j$ indicates the $g-r$, $i-z$, and $g-i$ colors, 
  $a_j$ and $b_j$ are the color boundaries of the complexes, and
  $\mu_j$ and $\sigma_j$ are the color and uncertainty of the SSO. Hence, 
  for a given observation, the volumes of all intersections sum to one.
  
\begin{figure*}[t]
    \centering
    \includegraphics[width=0.95\hsize]{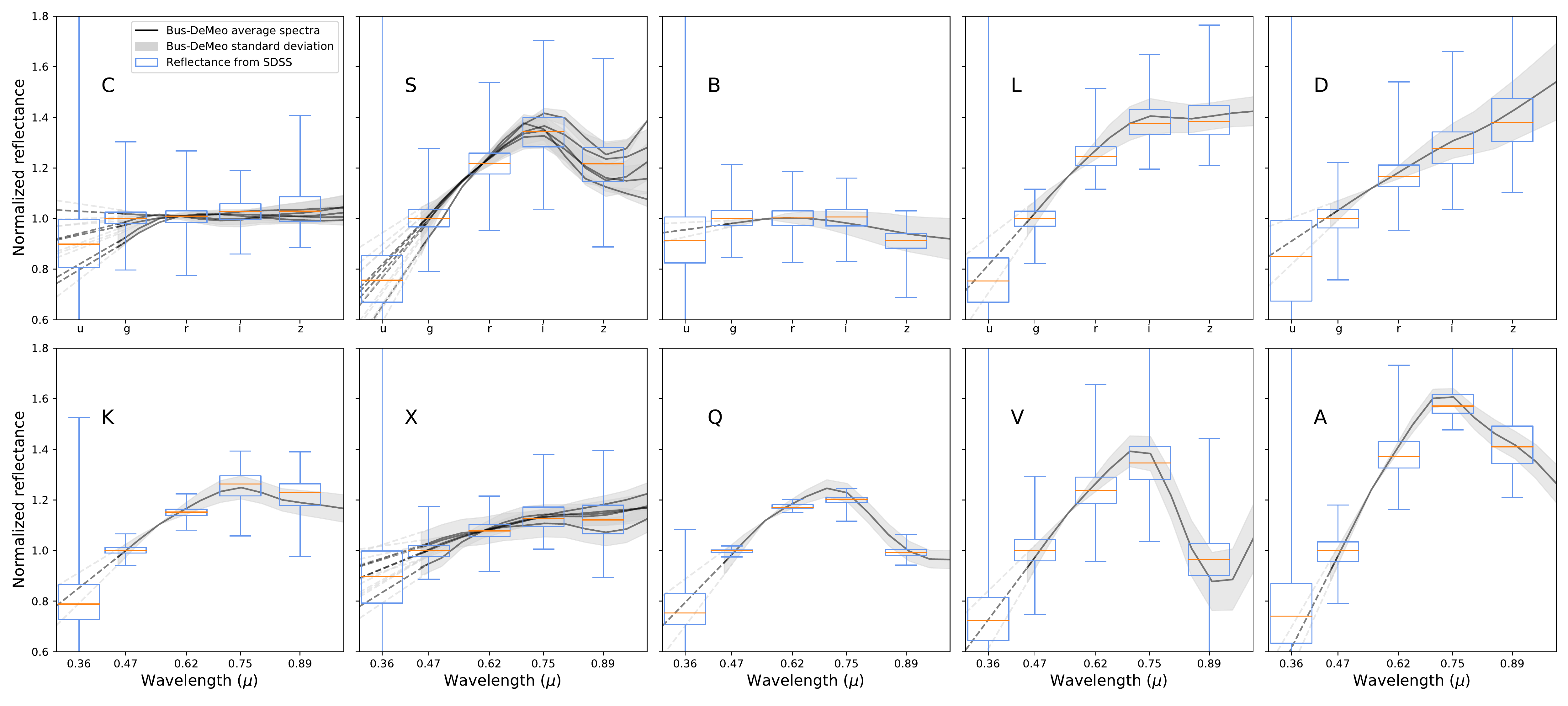}
    \caption{Pseudo reflectance spectra of asteroids observed
        by the SDSS, grouped by taxonomic class.
        The distribution of values for each band is
        represented by whiskers (95\% extrema, and the 25, 50, and 75\% quartiles).
        For each, we also represent
        the associated template spectra of the Bus-DeMeo taxonomy
        \citep{2009Icar..202..160D}.
        The Bus-DeMeo spectra do not cover the wavelength range covered by the
        \bandu filter, so we present a simple linear extrapolation.
        }
    \label{fig:spectra}
\end{figure*}  
  
 \paragraph{A probabilistic approach.}
  These normalized volumes correspond to the probabilities,
  $\mathcal{P}_k$, of pertaining
  to each taxonomic class.
  An observation fully fitting inside a given class cuboid
  will have a probability of 1 of being of that class. Conversely, an observation
  whose 3D Gaussian overlaps with two classes in a 1/3--2/3 proportion will have a 
  0.33 probability of being of the first class and a 0.66 probability of
  being of the second. 
  
  \paragraph{Handling of multiple observations.}
  The probabilistic approach offers a straightforward solution to multiple
  observations of the same object. The total probability for each
  class ($\mathcal{P}_{k}$)
  is the sum of the probabilities for that class over all observations ($j$), 
  weighted by the uncertainty of each observation ($V_{\sigma, j}$), where:
  
  \begin{equation}
      w_j = \frac{\sum{V_{\sigma, j}^{-2}}}{V_{\sigma, j}^{-2}}
  .\end{equation}

  \paragraph{Assignation of a class to each asteroid.}
  We assigned to each object its most probable class as derived from all its observations.
  The sole exception is the unknown class (U), which is only assigned if its probability is strictly
  equal to 1.
  Otherwise, whenever U is the most probable
  (due to poor photometric accuracy for instance),
  we assigned the second most probable class.
  
  We report in the catalog the probabilities for each class for each observation,
  together with the most probable class for that observation. We also list the 
  uncertainty, $V_{\sigma}$, and the most probable class after the combination of all the
  observations of a given SSO. 
  
  We present in \Autoref{fig:spectra} the pseudo reflectance spectra
  of each class, computed using the Solar colors of \citet{2006MNRAS.367..449H},
  compared with the template spectra of the associated Bus-DeMeo classes.
  The correspondence of the median SDSS spectra with the template spectra
  validates the approach described above.
  The spread of values is larger for the SDSS, but the sample size is  four orders
  of magnitude larger than the 371 spectra that define the Bus-DeMeo
  taxonomy.
  
\begin{figure*}[ht]
    \centering
    \includegraphics[width=0.99\hsize]{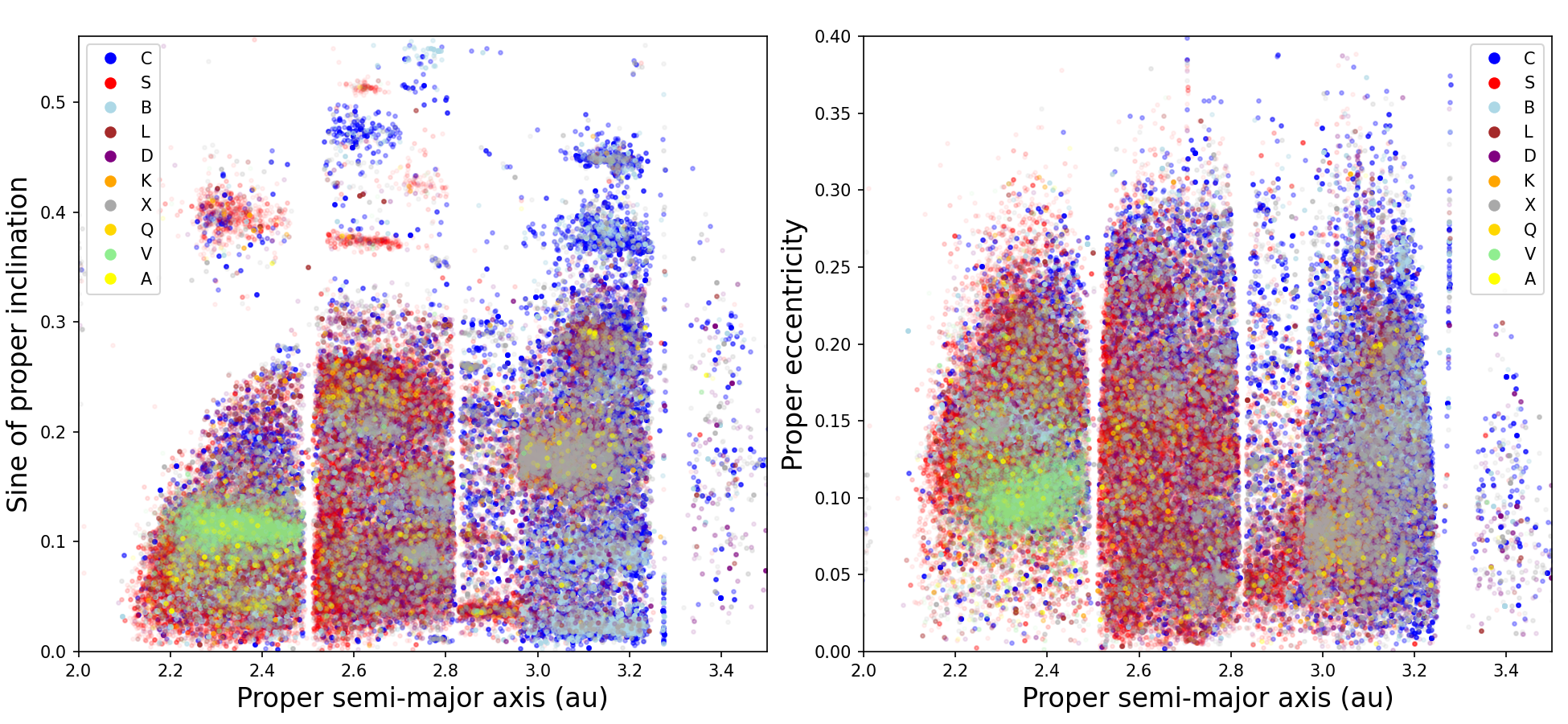}
    \caption{Orbital distribution of the SSOs reported here, color-coded by taxonomic class. }
    \label{fig:regions}
\end{figure*}

  We also illustrate the catalog by presenting in \Autoref{fig:regions}
  the orbital distribution of taxonomic classes.
  The general trend of S-types dominating the inner belt and 
  C-types the outer belt is clearly visible. The dynamical families are also
  easy to identify.

\begin{figure}[ht]
     \centering
     \includegraphics[width=1.0\hsize]{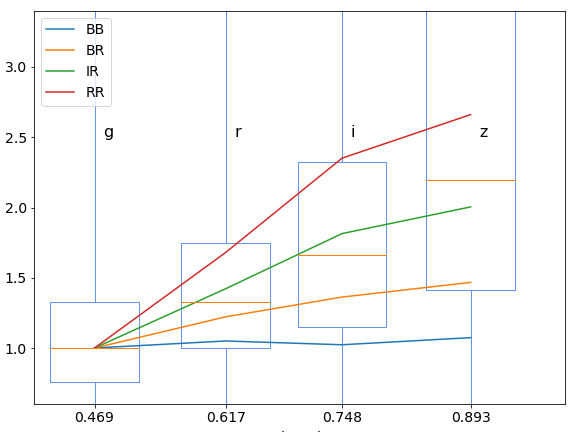}
     \caption{Pseudo reflectance spectra of KBOs from the SDSS.
        The BB (blue), BR (intermediate blue-red) , IR (moderately red), and RR (red)
        references are the median B-V, V-R, and R-I
        colors from \citet{2008-SSBN-3-Fulchignoni}, converted into
        SDSS colors.
        }
     \label{fig:spectra_oss}
 \end{figure}

  The Bus-DeMeo classification scheme we applied is adapted for asteroids but not
  for comets nor KBOs, which generally display featureless spectra in the visible,
  with spectral slopes ranging from X-type to D-type and even redder
  \citep{2006-MNRAS-373-Snodgrass, 2008-SSBN-3-Fulchignoni}.
  We did not apply a different scheme to comets and KBOs, but we present their
  pseudo reflectance spectra in \Autoref{fig:spectra_oss} compared with
  template spectral classes.

\section{Conclusions\label{sec:conclusion}}

  \indent Based on a new extraction of moving objects in the SDSS, 
  and following a suite of filters to minimize contamination, 
  we release a catalog of
  \numb{\ssotot} entries, consisting of
  \numb{\ssoknown} observations of
  \numb{\ssouniq} known and unique SSOs together with
  \numb{\ssounknown} observations of moving sources not linked with any known SSO.
  The catalog contains the 
  SDSS identification,
  astrometry,
  photometry, 
  SSO identification,
  geometry of observation, 
  taxonomy, and
  quality flag for each observation.
  Its content is fully described in \Autoref{app:cat}.

  \indent The catalog completeness is estimated to 
  be about \numb{95}\%
  and the purity to be above
  \numb{95}\%
  for known SSOs (see \Autoref{sec:purity} for details).
  The present catalog contains
  \numb{\ssoInADRfour} (\numb{85\%}) of
  the \numb{\nummoc} sources released in the ADR4 
  and
  \numb{93}\% of the NEAs and
  Mars-crossers from \citet{2016Icar..268..340C}.
  The missing sources are found to be 
  fixed-coordinate sources (identified by comparing
  their coordinates with the Pan-STARRS catalog using the procedure
  described in \Autoref{ssec:PS1}) in 
  \numb{\ADRfourInPSOne} cases.
  The remaining \numb{\ADRfourRemains} sources were located in fields
  marked as affected by poor weather conditions, which we excluded from the
  present analysis (\Autoref{sec:extract}).


\begin{acknowledgements}
  This  research has been conducted within the NEOROCKS project, which
  has received funding from the European Union's Horizon 2020 research
  and innovation programme under grant agreement No 870403. 
  
  \indent Funding for the SDSS and SDSS-II has been provided by the Alfred P. Sloan 
  Foundation, the Participating Institutions, the National Science Foundation, 
  the U.S. Department of Energy, the National Aeronautics and Space Administration, 
  the Japanese Monbukagakusho, the Max Planck Society, and the Higher Education 
  Funding Council for England. The SDSS Web Site is 
  \url{http://www.sdss.org}.\\

  \indent The SDSS is managed by the Astrophysical Research Consortium for the 
  Participating Institutions. The Participating Institutions are the American 
  Museum of Natural History, Astrophysical Institute Potsdam, University of 
  Basel, University of Cambridge, Case Western Reserve University, University of 
  Chicago, Drexel University, Fermilab, the Institute for Advanced Study, the 
  Japan Participation Group, Johns Hopkins University, the Joint Institute for 
  Nuclear Astrophysics, the Kavli Institute for Particle Astrophysics and Cosmology,
  the Korean Scientist Group, the Chinese Academy of Sciences (LAMOST), 
  Los Alamos National Laboratory, the Max-Planck-Institute for Astronomy (MPIA), 
  the Max-Planck-Institute for Astrophysics (MPA), 
  New Mexico State University, Ohio State University, University of Pittsburgh, 
  University of Portsmouth, Princeton University, the United States Naval 
  Observatory, and the University of Washington.\\

  \indent The Pan-STARRS1 Surveys (PS1) and the PS1 public science archive have been
  made possible through contributions by the Institute for Astronomy, the
  University of Hawaii, the Pan-STARRS Project Office, the Max-Planck Society
  and its participating institutes, the Max Planck Institute for Astronomy,
  Heidelberg and the Max Planck Institute for Extraterrestrial Physics,
  Garching, The Johns Hopkins University, Durham University, the University of
  Edinburgh, the Queen's University Belfast, the Harvard-Smithsonian Center
  for Astrophysics, the Las Cumbres Observatory Global Telescope Network
  Incorporated, the National Central University of Taiwan, the Space Telescope
  Science Institute, the National Aeronautics and Space Administration under
  Grant No. NNX08AR22G issued through the Planetary Science Division of the
  NASA Science Mission Directorate, the National Science Foundation Grant No.
  AST-1238877, the University of Maryland, Eotvos Lorand University (ELTE),
  the Los Alamos National Laboratory, and the Gordon and Betty Moore
  Foundation.\\

  \indent This research made use of the cross-match service provided by CDS, Strasbourg 
  \citep{2017AA...597A..89P},
  the IMCCE's \skybot and Skybot3D VO tools
  \citep{2006-ASPC-351-Berthier, 2016-MNRAS-458-Berthier},
  and TOPCAT/STILTS
  \citep{2005ASPC..347...29T}.
  Thanks to the developpers.

\end{acknowledgements}

\bibliographystyle{aa} 
\bibliography{40430corr} 

\begin{appendix} 

\section{Description of the SSO catalog\label{app:cat}}

\newcounter{rowcount}
\setcounter{rowcount}{-2}

\bottomcaption{Description of the SDSS SSO catalog columns. The catalog is available in \href{https://cds.u-strasbg.fr/}{Centre de Données astronomiques de Strasbourg (CDS)}.}
\tabletail{\hline}

\begin{xtabular}[H]{|@{\stepcounter{rowcount}
\ifnum\value{rowcount}=0{\makebox[1.5em][c]{\textbf{ID}}}
\else
        \makebox[1.5em][r]{\therowcount\xspace}
\fi}
|m{18mm}|c| m{4cm}|}
    \hline
    \textbf{Name} & \textbf{Unit} & \textbf{Description} \\ \hline
    objID &  & Unique SDSS identifier \\ \hline
    run &  & SDSS Run number \\ \hline
    camcol &  & SDSS Camcol number \\ \hline
    field &  & SDSS Field number \\ \hline
    ra & deg & J2000 Right Ascension (r-band) \\ \hline
    dec & deg & J2000 Declination (r-band) \\ \hline 
    tai\_r & day & MJD time of observation (TAI) in r filter \\ \hline  
    jd & day &  Julian time (UTC) \\ \hline \hline
    
    rowv & arcsec/hr & Row-component of object's velocity \\ \hline 
    rowvErr & arcsec/hr & Row-component of object's velocity error \\ \hline 
    colv & arcsec/hr & Column-component of object's velocity \\ \hline
    colvErr & arcsec/hr & Column-component of object's velocity  error\\ \hline
    vel & arcsec/hr & Object's velocity  \\ \hline
    velErr & arcsec/hr & Object's velocity error\\ \hline
    
    type\_r &  & SDSS Type classification of the object (3 - galaxy, 6 - star) \\ \hline \hline
    
    psfMag\_u & mag & \multirow{4}{*}{\parbox{4cm}{PSF magnitude in each filter}} \\ \cline{1-2}
    psfMag\_g & mag &  \\ \cline{1-2}
    psfMag\_r & mag &  \\ \cline{1-2}
    psfMag\_i & mag &  \\ \cline{1-2}
    psfMag\_z & mag &  \\ \hline \hline
    psfMagErr\_u & mag & \multirow{4}{*}{\parbox{4cm}{PSF magnitude error in each filter}} \\ \cline{1-2}
    psfMagErr\_g & mag &  \\ \cline{1-2}
    psfMagErr\_r & mag &  \\ \cline{1-2}
    psfMagErr\_i & mag &  \\ \cline{1-2}
    psfMagErr\_z & mag &  \\ \hline \hline
    
    offsetRa\_u & arcsec & \multirow{10}{*}{\parbox{4cm}{Filter positions RA and Dec minus final coordinates}} \\* \cline{1-2}
    offsetDec\_u & arcsec & \\ \cline{1-2}
    offsetRa\_g & arcsec & \\ \cline{1-2}
    offsetDec\_g & arcsec & \\ \cline{1-2}
    offsetRa\_r & arcsec & \\ \cline{1-2}
    offsetDec\_r & arcsec & \\ \cline{1-2}
    offsetRa\_i & arcsec & \\ \cline{1-2}
    offsetDec\_i & arcsec & \\ \cline{1-2}
    offsetRa\_z & arcsec & \\ \cline{1-2}
    offsetDec\_z & arcsec & \\ \hline \hline 
    \shrinkheight{-0.5in}

    Number & & Solar System object number \\ \hline
    Name & & Solar System object name \\ \hline
    ra\_sb & deg & SkyBot predicted RA coordinate \\ \hline
    dec\_sb & deg & SkyBot predicted Dec coordinate \\ \hline
    dynclass & & SkyBot object dynamic class \\ \hline
    V & mag & SkyBot predicted visual magnitude \\ \hline
    posunc & arcsec & Positional uncertainty \\ \hline
    centerdist & arcsec & Angular distance of target from cone center \\ \hline 
    ra\_sb\_rate & arcsec/hr & RA rate of motion \\ \hline 
    dec\_sb\_rate & arcsec/hr & Declination rate of motion \\ \hline 
    geodist & au & Geocentric distance of target \\ \hline
    heliodist & au & Heliocentric distance of target \\ \hline
    alpha & deg & Solar phase angle \\ \hline
    elong & deg & Solar elongation angle \\ \hline
    \hline
    boffset & & Offset flag \\ \hline
    bphot\_u & & \multirow{5}{*}{\parbox{4cm}{Photometry flags}} \\ \cline{1-2}
    bphot\_g & & \\ \cline{1-2}
    bphot\_r & & \\ \cline{1-2}
    bphot\_i & & \\ \cline{1-2}
    bphot\_z & & \\ \hline
    bcolor\_ru & & \multirow{4}{*}{\parbox{4cm}{Color flags}} \\ \cline{1-2}
    bcolor\_rg & & \\ \cline{1-2}
    bcolor\_ri & & \\ \cline{1-2}
    bcolor\_rz & & \\ \hline
    bastrom\_i & & \multirow{3}{*}{\parbox{4cm}{Astrometry flags}} \\ \cline{1-2}
    bastrom\_u & & \\ \cline{1-2}
    bastrom\_z & & \\ \hline
    bknown & & Known SSO \\ \hline
     
    R2 & & Coefficient of determination \\ \hline 
    \hline
    Volume & & Color uncertainty \\ \hline
    p[C, S, B, L, D, 
    K, X, V, A, U] & & probability of complex values \\ \hline
    complex & & Most probably complex of individual observation \\ \hline
    pcomplex & & Complex probably value of individual observation\\ \hline
    gr\_complex & & Most probably complex of SSO \\ \hline
    gr\_pcomplex & & Complex probably value of SSO\\ \hline
    
    proper\_H & mag & Absolute magnitude (from astdys)\\ \hline
    proper\_a & au & Proper semimajor axis (from astdys) \\ \hline
    proper\_e &  & Proper eccentricity (from astdys)\\ \hline
    proper\_sinI &  & Sine of proper inclination (from astdys) \\ \hline
    
    osc\_H & mag & Absolute magnitude (from astorb) \\ \hline
    osc\_a & au & Semimajor axis  (from astorb) \\ \hline
    osc\_inc &  & Eccentricity (from astorb) \\ \hline
    osc\_e & deg & Inclination (from astorb) \\ \hline
\end{xtabular}

\newpage

\section{SQL code for moving object extraction\label{app:sql}}


\begin{lstlisting}[language={SQL}, caption={SDSS request.}, label={list:sql_sdss}]
FROM dbo.PhotoObjAll as p
WHERE 
  p.flags & dbo.fPhotoFlags('DEBLENDED_AS_MOVING') > 0
  AND p.flags & dbo.fPhotoFlags('SATURATED') = 0
  AND p.flags &  dbo.fPhotoFlags('BRIGHT') = 0 
  AND p.flags &  dbo.fPhotoFlags('EDGE') = 0
  AND (p.type_r = 6 OR p.type_r = 3) 
  AND power(p.rowv,2) + power(p.colv, 2) > 0.0025
  AND p.petroMag_r BETWEEN 14.5 AND 22.2
\end{lstlisting}

\begin{lstlisting}[language={SQL}, caption={GAIA request.}, label={list:sql_gaia}]
SELECT source_id, ra, dec, coord1(prop) AS ra_2004, coord2(prop) AS dec_2004, pmra, pmdec, phot_g_mean_mag
FROM (SELECT gaia.source_id, ra, dec, pmra, pmdec, phot_g_mean_mag,
            EPOCH_PROP_POS(ra, dec, parallax, pmra, pmdec, 0, ref_epoch, 2004) AS prop
      FROM gaiadr2.gaia_source AS gaia
            phot_g_mean_mag>=14 AND
            sqrt(power(gaia.pmra, 2) + power(gaia.pmdec, 2)) > 10
     ) AS subquery
\end{lstlisting}

\end{appendix}

\end{document}